\documentclass{article}

\usepackage[english]{babel}
\usepackage{float}
\usepackage{setspace} 
\usepackage[letterpaper,top=2cm,bottom=2cm,left=3cm,right=3cm,marginparwidth=1.75cm]{geometry}
\date{}
\usepackage{amsmath}
\usepackage{graphicx}
\usepackage[colorlinks=true, allcolors=blue]{hyperref}
\usepackage{authblk}

\title{MEV in fixed gas price blockchains: Terra Classic as a case of study}
\author[1]{Facundo Carrillo}
\author[2]{Elaine Hu}
\affil[1]{CONICET-Universidad de Buenos Aires, Instituto de Investigación en Ciencias de la Computación (ICC), Buenos Aires, Argentina}
\affil[2]{Flashbots}
\date{}                    
\begin{document}
\maketitle

\begin{abstract}
Maximum extractable value (MEV) has been extensively studied. In most papers, the researchers have worked with the Ethereum blockchain almost exclusively. Even though, Ethereum and other blockchains have dynamic gas prices this is not the case for all blockchains; many of them have fixed gas prices. Extending the research to other blockchains with fixed gas price could broaden the scope of the existing studies on MEV. To our knowledge, there is not a vast understanding of MEV in fixed gas price blockchains. Therefore, we propose to study Terra Classic as an example to understand how MEV activities affect blockchains with fixed gas price. We first analysed the data from Terra Classic before the UST de-peg event in May 2022 and described the nature of the exploited arbitrage opportunities. We found more than 188K successful arbitrages, and most of them used UST as the initial token. The capital to perform the arbitrage was less than 1K UST in 50\% of the cases, and 80\% of the arbitrages had less than four swaps. Then, we explored the characteristics that attribute to higher MEV.  We found that searchers who use more complex mechanisms, i.e. different contracts and accounts, made higher profits. Finally, we concluded that the most profitable searchers used a strategy of running bots in a multi-instance environment, i.e. running bots with different virtual machines. We measured the importance of the geographic distribution of the virtual machines that run the bots. We found that having good geographic coverage makes the difference between winning or losing the arbitrage opportunities. That is because, unlike MEV extraction in Ethereum, bots in fixed gas price blockchains are not battling a gas war; they are fighting in a latency war.
\end{abstract}

\section{Introduction}

MEV has been extensively studied, from mathematical models to data analysis applications \cite{qin2022quantifying, weintraub2022flash, flashbotsfrontrunning2022,daian2020flash}. In most studies, researchers have worked with the Ethereum blockchain almost exclusively, so extending the studies to other blockchains could help broaden the scope of MEV research. Understandably, one of the most important aspects of MEV is gas bidding.

In Ethereum, the miners include transactions from the mempool and propose a block with a subset of transactions in an arbitrary order. However, this arbitrary order is not random; many miners use gas to sort transactions and create new blocks since this strategy maximises their rewards. This incentive increases the gas price to a point that makes the network almost impossible to use for normal users.

Although this incentive is common in Ethereum, it is not present in all blockchains. Many blockchains based on Tendermint don't take dynamic gas prices into account to prioritise transactions. They use a fixed gas price and process transactions in the same order in which the validators receive them \footnote{ \href{https://docs.tendermint.com/v0.34/tendermint-core/mempool.html}{https://docs.tendermint.com/v0.34/tendermint-core/mempool.html} }. This difference could heavily impact the role and behaviour of MEV searchers. In blockchains with fixed gas prices, the only resource that searchers have is to recognize an MEV opportunity before other actors and respond quickly. This contrast could have enormous repercussions on the opportunities that MEV searchers choose to pursue. At first, there would be no possibility, by non-privileged users, of positioning one transaction ahead of another once the first one was already broadcasted. This restriction would forbid extracting value based on front-running and sandwich strategies, leaving only back-running as a resource for searchers. Even though searchers only have this strategy, arbitrages could be performed. 

Arbitrage is a particular way to extract value. It is a concept broadly used in traditional finances that has presented a great opportunity in blockchains, particularly since decentralised exchanges have become so popular. Arbitrage consists of taking advantage of a difference in prices in two or more markets. Users can buy an asset in one market at price A and sell it in another market at price B. This action results in a profit equals to B minus A. This practice is well established in many economic systems and allows different markets to stay synchronised between them. Many studies have presented profits higher than hundred of millions of dollars by applying arbitrage in Ethereum\cite{wang2022cyclic,makarov2020trading,jin2022detecting,hansson2022arbitrage,zhou2021just}.

The community understands the importance of this practice. Every day more data allows us to follow the progress in value extraction in different blockchains. For example, the Flashbots community offers a dashboard that measures MEV extraction on Ethereum in real-time\footnote{\href{https://explore.flashbots.net/}{https://explore.flashbots.net/}}, Skip Protocol provides data for Osmosis \footnote{\href{https://satellite.skip.money/}{https://satellite.skip.money/}}, and Marlin.Org for Polygon \footnote{\href{https://explore.marlin.org/}{https://explore.marlin.org/}}, among others. Although all these data allow us to quantify the amount of MEV extracted, they do not provide the full picture. There isn't a complete understanding of the type of extraction, the type of pairs exploited, and cycles, among others. In particular, those sites do not give enough information about searchers' behaviour and how they maximise profit.

We believe blockchains that don't use gas to prioritise transactions present MEV opportunities with their own dynamics. As we have previously mentioned, even though MEV has been extensively studied  (market designs and their implications \cite{whyyourblockchianneedsandmevsolution}, MEV in other blockchains \cite{abriefsurveyofmev}, among others), to our knowledge, there is not a full understanding of MEV in blockchains with fixed gas price mechanisms. Therefore, we propose the following general objective: To study the MEV opportunities in a blockchain with fixed gas price. To contribute to this general object, we propose to study the Terra Classic blockchain during the period from September 2021 until the de-peg event in May 2022, with the following specific objectives:

\begin{itemize}
\item Specific Objective 1: Study the characteristics of arbitrages in Terra Classic in the defined period. 
\item Specific Objective 2: Understanding which strategies perform searchers to improve profit.
\item Specific Objective 3: Understanding the timing-related characteristics of arbitrage.
\item Specific Objective 4: Create a dashboard to help with the analysis of this data 
\end{itemize}

\section{Methods}
\subsection{Dataset}

Terra Classic suffered a particular and non-typical behaviour during and after the de-peg event in May 2022. Due to this phenomenon, we decided to use the last version of Terra Classic (Columbus-5) before the de-peg event. This dataset contains 2,801,944 blocks, from block 4734841 (2021-10-01 08:31:52 UTC) to block 7536784 (2022-05-07 00:00:04 UTC).

In order to get the data, we ran a Terra Classic node using the snapshot available in ChainLayer \footnote{\href{https://quicksync.io/networks/terra.html}{https://quicksync.io/networks/terra.html}}. This website offers different blockchain snapshots to speed up synchronisation. We used the Terra Classic's code node available on its repository \footnote{\href{https://github.com/terra-money/classic-core}{https://github.com/terra-money/classic-core}}.

After we set up the node, we iterated over all the blocks using Terra Classic's API to consume data. For every block, we used the Tendermint RPC to get the block with its transactions coded in base 64. Then we also used the Terra RPC to get the block result. This entry returns the logs of every transaction after the block is proposed. Having this information is very useful as it helps us understand if the transactions succeeded and characterise its nature. Information on these two APIs is available in the official documentation \footnote{\href{https://classic-docs.terra.money/docs/develop/endpoints.html}{https://classic-docs.terra.money/docs/develop/endpoints.html}}.

\subsection{Arbitrage identification}

We created an algorithm to understand if a successful transaction performed an arbitrage. To do this, we used the transaction logs. Terra transactions' logs have a lot of information. Of all the information available, we used only the wasm\cite{haas2017bringing} log to understand if a transaction performed an arbitrage. For this, we define a function that identifies a series of actions. These actions try to represent a swap between tokens. Typically, these actions are simple swaps, but sometimes they are more difficult to understand, for example, burning a token and minting a new one. For this purpose, we define an action with the following fields: pair address, token in, token out, amount in and amount out. A transaction then contains a list of actions.  With all the information, we define an entire arbitrage as a list of these actions with the following constraints:

\begin{enumerate}
    \item Successive tokens match

\begin{equation}
\forall i \in [0 ... |actions|) \ actions[i].token_{out} = actions[i+1].token_{in}  \ 
\end{equation}

Note that $actions[i]$ represents the i-th action, $actions[i+1]$ represents the following action, and $actions[i-1]$ represents the previous action.

\item Arbitrage generates profit 

\begin{equation}
    actions[ |actions| - 1].amount_{out}  > actions[0].amount_{in}
\end{equation}

Note that fees are not taken into account.

\item Arbitrage starts and ends with the same token
\begin{equation}
    actions[|actions| -1 ].token_{out} = actions[0].token_{in}
\end{equation}

\end{enumerate}

\begin{figure}[H]
\centering
\includegraphics[width=1\textwidth]{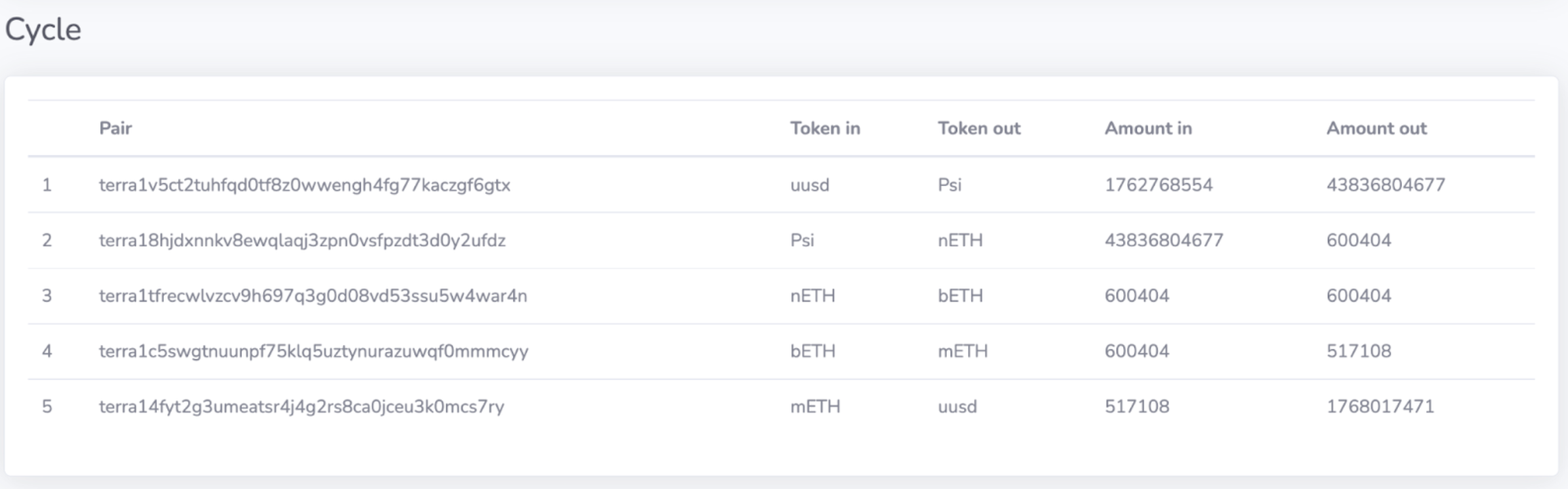}
\caption{\label{fig:fig17_method}
Example of how the platform shows a list of actions (\href{https://facuzeta.github.io/frp/dashboard/8297c618c35e363edc2030e165eca9c01d7a66eb4736a9b0bf255d5d6ec705ce}{link to dashboard}). }
\end{figure}

For example, Figure \ref{fig:fig17_method} shows the actions list as a table for a particular transaction.

If we can parse the logs and create a consistent list of actions that complies with the restrictions, we flag this transaction as a successful arbitrage. In summary, we deem an arbitrage successful if it generates profit (without taking into account fees); and the transaction is included in a block; and its execution ends without being reverted. The code for parsing logs and creating the list of actions are available in the repository (see \href{sec:data-code-and-notebooks}{Data, Code and Analysis notebooks})

\subsubsection{Inference of arbitrage in failed transactions}
\label{sec:inference-of-arbitrage-in-failed-transactions}

Transactions revert for different reasons. In most cases, lack of profit is the cause. When transactions revert, they don't produce the same wasm logs we use to understand if the transactions are arbitrage. Therefore, we developed a heuristic to flag arbitrage attempts. We define a transaction as failed transaction if:

\begin{enumerate}
    \item The sender address has signed at least one successful arbitrage
    \item The contract address has executed at least one successful arbitrage
    \item The execute-message resembles another one produced by the same contract in at least one successful arbitrage.
\end{enumerate}

\subsection{Dashboard}
An interactive dashboard is developed using Django \cite{forcier2008python} to display the charts and findings. The dashboard is available at: \href{https://facuzeta.github.io/frp/dashboard}{facuzeta.github.io/frp/dashboard}. 

\subsubsection{Time Analysis}
\label{subsubsection:time-analysis}
In order to analyse the effect of latency, we need to record transactions in different geographic locations. For this, we designed an experiment in which we deployed 84 instances well distributed around the world using AWS cloud services. We created instances in the following regions: US East (Ohio), US East (N. Virginia), US West (N. California), US West (Oregon), Africa (Cape Town), Asia Pacific (Hong Kong), Asia Pacific (Jakarta), Asia Pacific (Mumbai), Asia Pacific (Osaka), Asia Pacific (Seoul), Asia Pacific (Singapore), Asia Pacific (Sydney), Asia Pacific (Tokyo), Canada (Central), Europe (Frankfurt), Europe (Ireland), Europe (London), Europe (Milan), Europe (Paris), Europe (Stockholm), Europe (Zurich), Middle East (Bahrain), Middle East (United Arab Emirates), South America (São Paulo). 

After we deployed the instances, we recorded the transactions that arrived at the mempool. Every node started with the same list of peers to connect (the default peer list offered by the official documentation). But these peers are not necessarily connected with the same set of nodes throughout the experiment. This is due to the fact that Tendermint P2P layer connects, disconnects, and asks for new peers depending on different statistics. Besides the default configuration, we added two constraints : 1) our 84 nodes are not connected to each other, and 2) Our nodes do not broadcast transactions from their mempool. 

Finally, we measured the operation system time differences among instances to make sure that they had an error of less than 15 milliseconds.

\subsection{Data, Code and Analysis notebooks}
\label{data-code-and-notebooks}
All the data and code used in this work are available in the following repositories: 

\begin{itemize}
    \item \href{https://github.com/facuzeta/frp-mev-fixed-gas-price-dashboard}{https://github.com/facuzeta/frp-mev-fixed-gas-price-dashboard}
    \item \href{https://github.com/facuzeta/frp-mev-fixed-gas-price-analysis}{https://github.com/facuzeta/frp-mev-fixed-gas-price-analysis }
    
\end{itemize}

\section{Results}
Blocks in Terra contain different types of transactions. Among them are: transactions to initialise a contract, transactions to send tokens, transactions to vote on governance decisions, and transactions to execute smart contracts, and others. Arbitrages are usually carried out through transactions that execute smart contract functions. Taking this into account, we only use Execute-Msg type transactions to analyse the data.

\begin{figure}[H]
\centering
\includegraphics[width=0.5\textwidth]{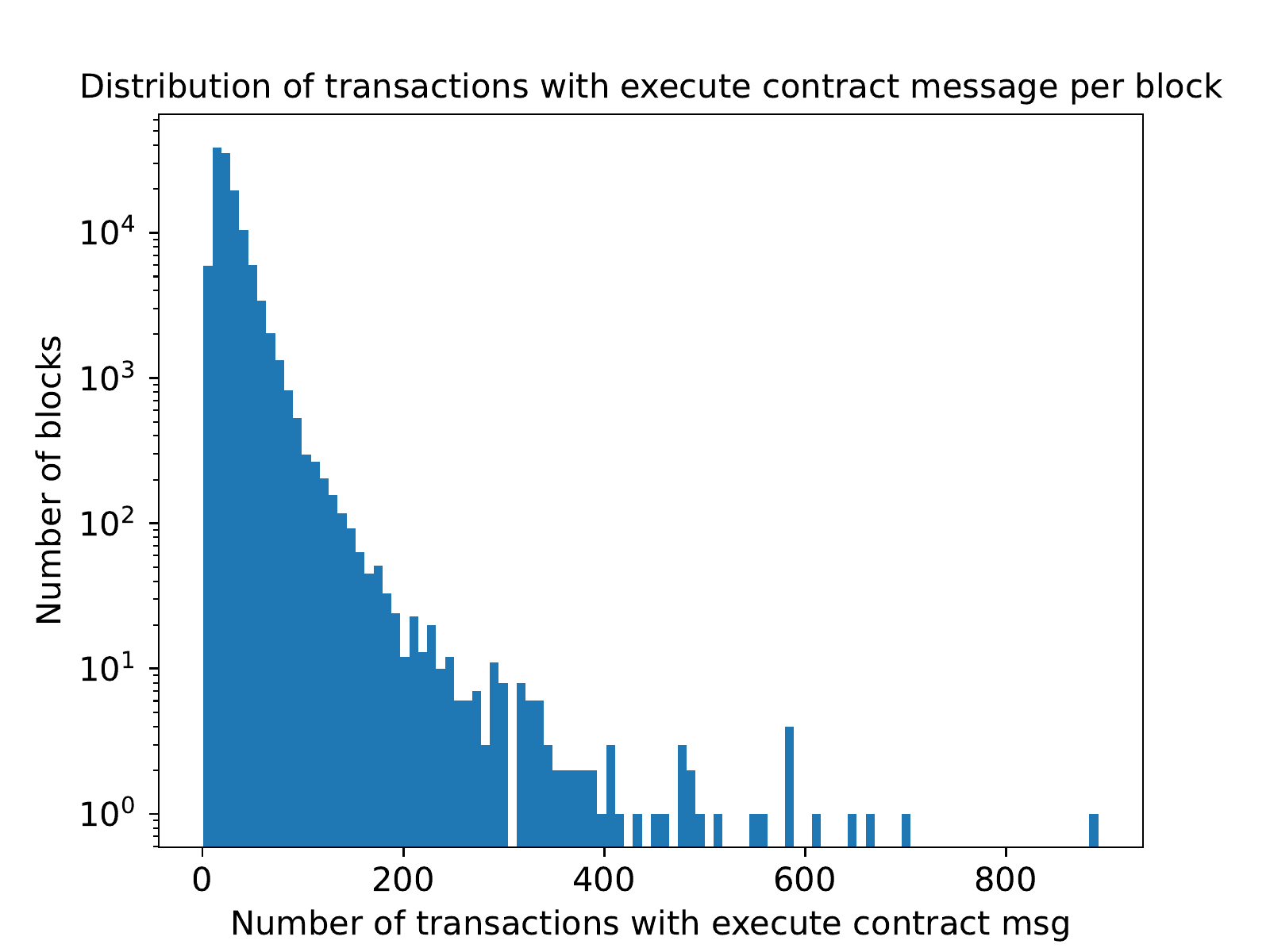}
\caption{\label{fig:fig1_distribution_of_transactions_with_execute_contract_message_per_block}Distribution of transactions with execute contract message per block.}
\end{figure}

Our dataset contains 2,801,944 blocks, from block 4734841 2021-10-01 08:31:52 UTC to block 7536784 2022-05-07 00:00:04 UTC. These blocks contain more than 117M transactions of any type and more than 37M execute-message transactions. The number of transactions in blocks varies depending on the transactions available in the mempool at the time of the block creation. On average, there are 28.10$\pm$ 22 (mean and standard deviation) transactions per block with a median of 23. Even though there are blocks with more than 800 transactions, 90\% of blocks contain less than 50 transactions. Figure \ref{fig:fig1_distribution_of_transactions_with_execute_contract_message_per_block} shows the distribution of the number of transactions with execute contract message per block.

\subsection{Successful arbitrages}
We found 188,564 arbitrage that were mined by the validators. These transactions are distributed in many blocks. However, only 4.48\% (125K) of the blocks presented at least one successful arbitrage.

\begin{figure}[H]
\centering
\includegraphics[width=0.9\textwidth]{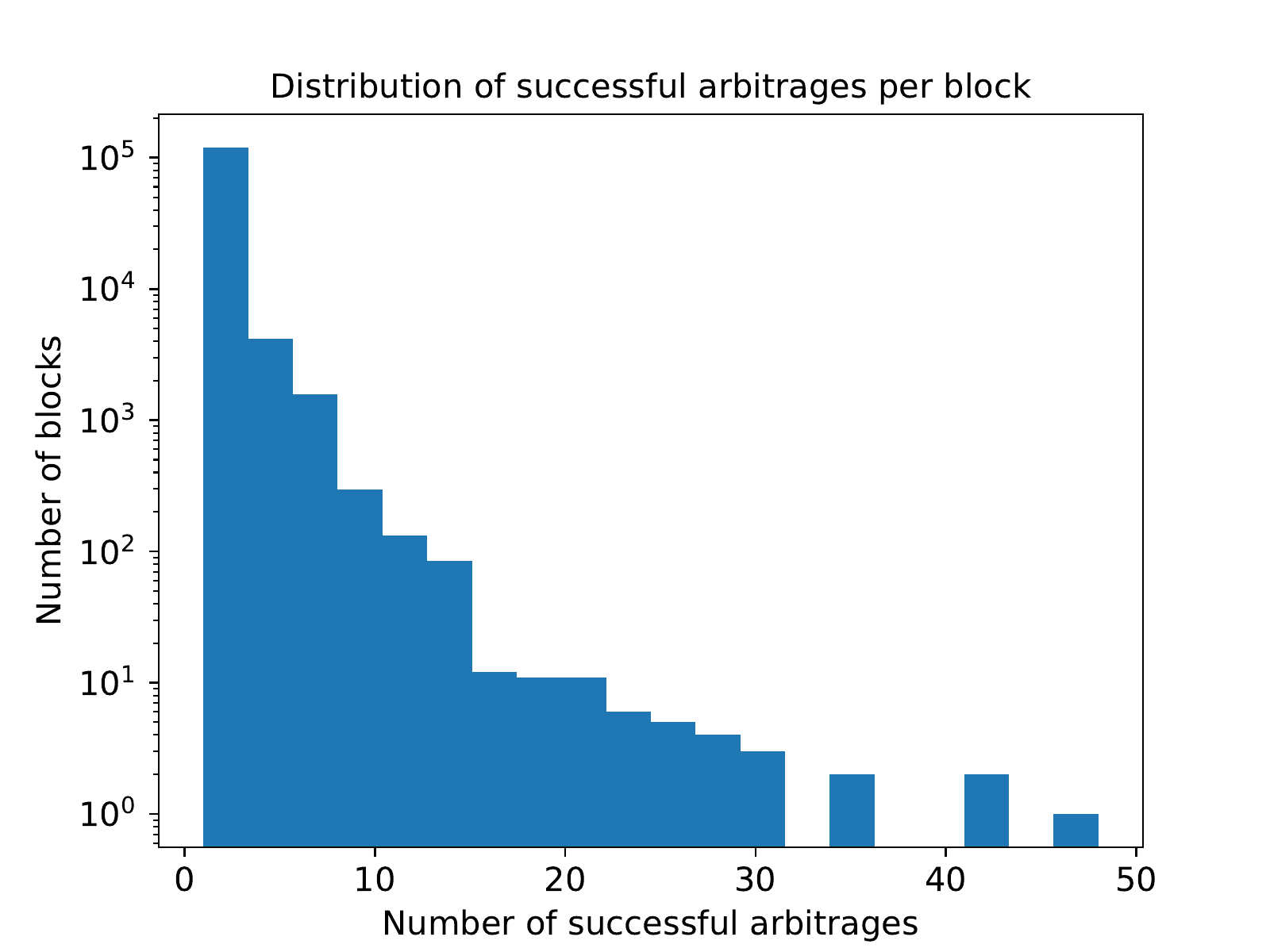}
\caption{\label{fig:fig2_distribution_of_succesful_arbs_per_block}Distribution of successful arbitrages per block  with at least one successful arbitrage.}
\end{figure}

In the blocks where there is at least one arbitrage, in 73\% of the cases, only one arbitrage occurs. In 16\% of the blocks, two successful arbitrages happen, and in 5\% of the cases, three successful arbitrages occur. That is, in 94\% of the blocks with successful arbitrages, a maximum of 3 arbitrages per block occur. The Figure \ref{fig:fig2_distribution_of_succesful_arbs_per_block} illustrates the distribution of the number of arbitrages in the blocks with at least one successful arbitrage. This variability is probably due to different arbitrage opportunities occurring on certain blocks. 

\subsection{
Identifying searchers
}
Identifying searchers is a difficult task because of the on-chain data availability. We can only see addresses signing transactions to execute smart contracts. In the 188K arbitrages, we found 517 different sender addresses and 167 different smart contracts.

Grouping the arbitrages by the sender, only the top 10 generate 40\% of the successful arbitrages. (see Table \ref{tab:tab1_top10_arb_sender}). Grouping by the smart contract, the top 10 contracts used by searchers account for 49\% of successful arbitrages (see Table \ref{tab:tab2_top10_arb_contract}).

\begin{table}
\centering
\begin{tabular}{|l|l|l|l|}
\hline
Sender &  Successful arbitrages &  Contracts &  \% \\\hline
terra1fgef888tuj3g8tmpxp4klvthq4eqk8a2yw7vsw &                     27197 &                    6 &   14.42 \\\hline
terra1lywrs72lpsptu68usehlae5yq94wsuk2g6fa3m &                     10232 &                   13 &    5.42 \\\hline
terra1n3s52w97j599sq3k4wcr3j0027j0stnfttnjvg &                      7385 &                    7 &    3.91 \\\hline
terra1nyud8zzctrks6ltnhhn4q820umnvqg39hc6ydz &                      7193 &                    2 &    3.81 \\\hline
terra16kd4ucrkj3mu2xpt25075s3jsztkx9ltf3tj9y &                      4921 &                    1 &    2.60 \\\hline
terra1xjpy3hu7qzun8p3h3cuf4w2v4f98w376mjj74l &                      4797 &                   13 &    2.54 \\\hline
terra140d3kgus5nxnqextv6fahzncfvp90r7t0s6mjp &                      3684 &                    5 &    1.95 \\\hline
terra1lxyhvhhdjc3sxk0kjeljtqnxjf7ek6asxxyp7q &                      3645 &                    3 &    1.93 \\\hline
terra1cxspyydlp4qdu0pzetg9au549n29tnvs4t3kej &                      3334 &                    2 &    1.76 \\\hline
terra1qfg2exflsshg23gs5w4ml37hwnunzthtx2gpp6 &                      3288 &                    2 &    1.74 \\\hline

\end{tabular}
\caption{\label{tab:tab1_top10_arb_sender}Top 10 arbitrage senders}
\end{table}

\begin{table}
\centering
\begin{tabular}{|l|l|l|l|}
\hline
Contract &  Successful arbitrages &  Senders &  \% \\ \hline
terra1p9zmnwqrfpzx0k585r3lhrqhlakmrnar2cljt9 &                     17130 &                  1 &    9.08 \\\hline
terra1kz93zt3ng7kags09qy06a079wjl3qpzfp0axsn &                     14195 &                  8 &    7.52 \\\hline
terra1nuzl3sppu0suws9zml5psu8tmuqk5dnr88n8kc &                     10664 &                  8 &    5.65 \\\hline
terra10j88evvssl4m0ztj2hcelf993j5a0xd3ezd7se &                     10314 &                 80 &    5.46 \\\hline
terra1myl5pk2a7qj37yu6dgmy68rxkznmq9nrk69nrj &                      8367 &                 50 &    4.43 \\\hline
terra1wswy9763nhugvphchcdc438r3ejvskjjz5h0rs &                      7731 &                  8 &    4.09 \\\hline
terra139y02s9urkkyesukndrqdjmqj7gkk5dltd05v8 &                      7060 &                  8 &    3.74 \\\hline
terra19rltg2vaurffa25cvtm5zey9muhaxflfxt5e2p &                      6152 &                  1 &    3.26 \\\hline
terra1egqmcaupc87sdg6sqmzypuv06ykxkek3j7s4w5 &                      5927 &                  1 &    3.14 \\\hline
terra1sg877xutwmyvsz45jrlwcdhx07jac33433499e &                      5634 &                  3 &    2.98 \\\hline
\end{tabular}
\caption{\label{tab:tab2_top10_arb_contract}Top 10 arbitrage contracts}
\end{table}

A contract is not necessarily used by a single address and an address does not necessarily use a single smart contract to carry out arbitrages. In fact, only 55\% of the contracts are used by a single sender, while 61\% of the senders use a single contract.

Given this, we proposed a method to infer searchers. The method consists of grouping sender addresses and smart contracts related to each other. We defined that an address is related to a smart contract if the sender signed a transaction that executed the smart contract. Then, we extended this using a transitive property. 

To compute this, we created a graph where each node represented a sender address or a smart contract. Then, the two nodes had a common edge if there was a transaction calling the smart contract signed by the address. Next, we executed the analysis of connected components. Every connected component was a searcher, and the addresses and smart contracts belonged to it.

\begin{figure}[H]
\centering
\includegraphics[width=0.9\textwidth]{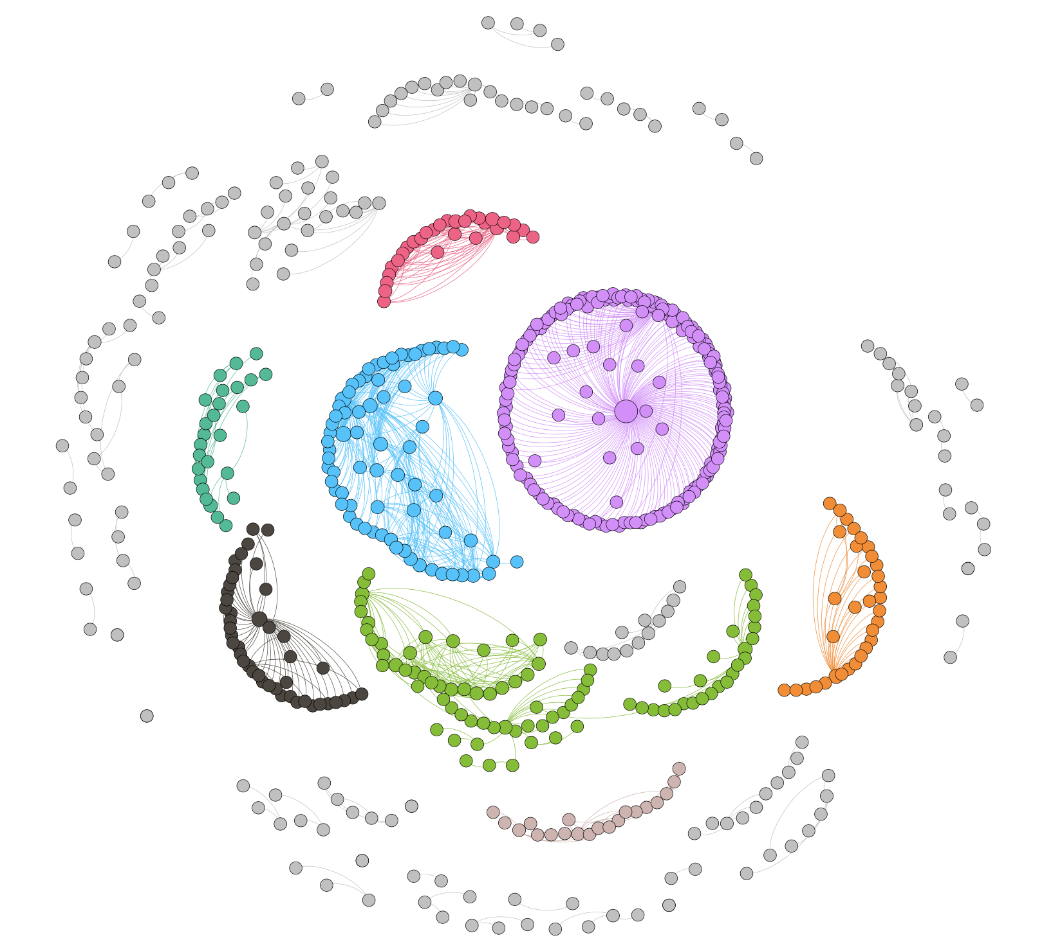}
\caption{\label{fig:fig3_graph} In this graph, each node represents a sender address or a smart contract. Nodes have a common edge if there is a transaction calling the smart contract and it is signed by the sender's address. The largest nine connected components are coloured. The rest of the connected components are grey. The size of the node represents the node degree. The distance between nodes does not add extra information.}
\end{figure}

Figure \ref{fig:fig3_graph} shows the graph and the inferred searchers.

\begin{figure}[H]
\centering
\includegraphics[width=1\textwidth]{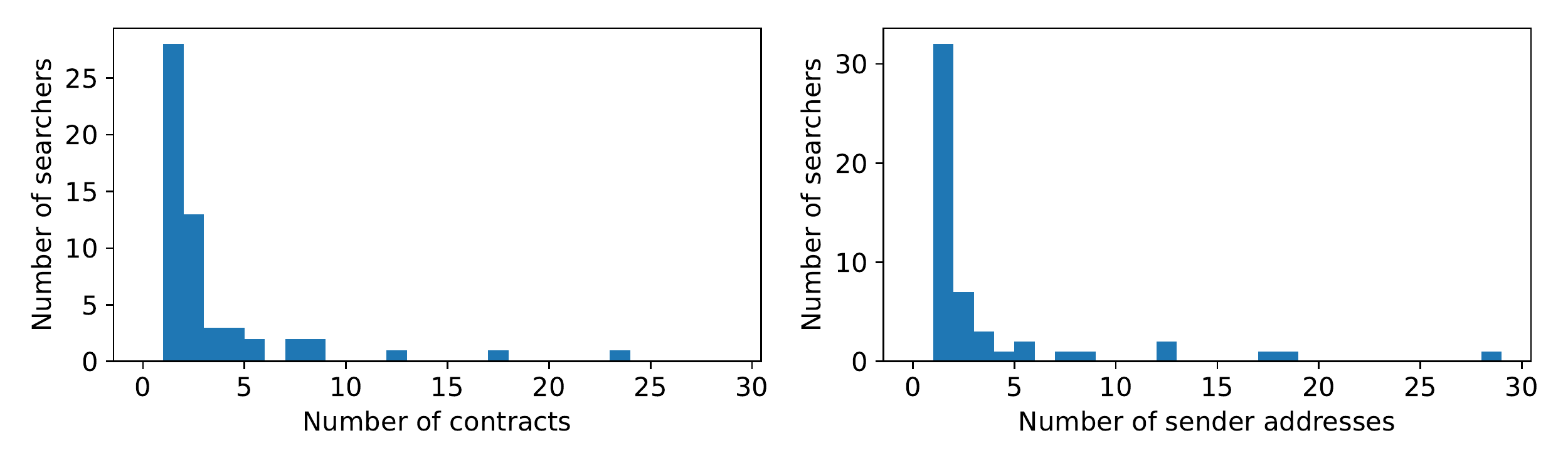}
\caption{\label{fig:fig4_number_of_contracts_and_senders_per_searcher} The left panel shows the histogram of the number of contracts per searcher. The right panel shows the distribution of the number of sender addresses per searcher. }
\end{figure}

Our method identified 56 connected components or searchers. 50\% of them used only one smart contract, and  57\% of the searchers used only one address to sign transactions. Figure \ref{fig:fig4_number_of_contracts_and_senders_per_searcher} shows the distribution of contracts and senders by searchers.

\begin{figure}[H]
\centering
\includegraphics[width=0.9\textwidth]{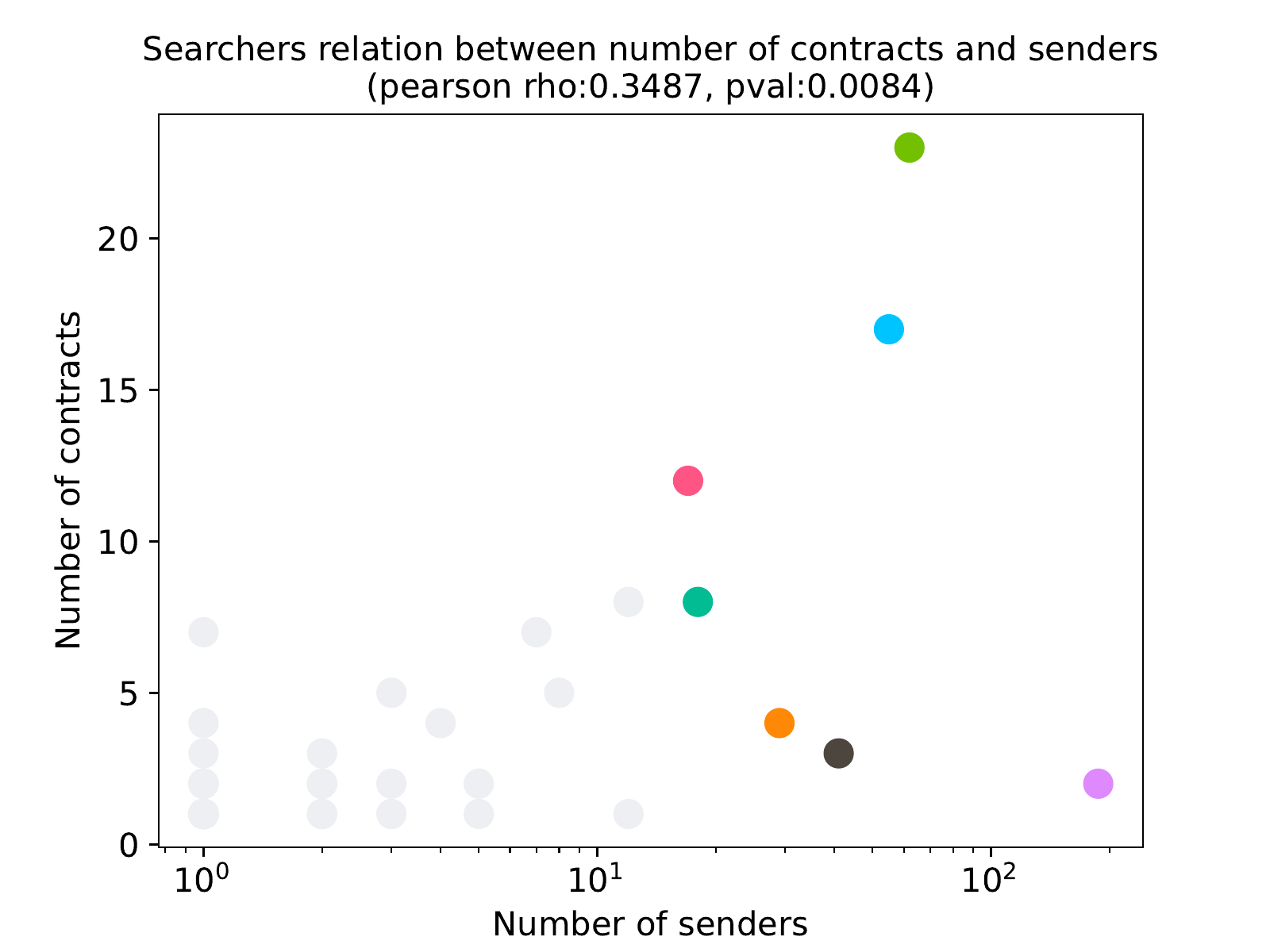}
\caption{\label{fig:fig5_searchers_relation_between_number_of_contracts_and_senders} Scatter plot of the number of different senders' addresses and contracts by searchers. The colour code follows the one presented in Figure \ref{fig:fig3_graph}. }
\end{figure}

There was a positive correlation (Pearson rho=0.3487, p-value=0.0084) between the number of contracts and addresses that searchers used. The number of contracts and addresses could describe the complexity of searchers' strategies. Some searchers used only one smart contract and one wallet to sign transactions. More sophisticated searchers could have multiple contracts and wallets or sender addresses. Figure \ref{fig:fig5_searchers_relation_between_number_of_contracts_and_senders} shows this correlation.

\subsection{Other characteristics of arbitrage
}
\subsubsection{Token In}
\begin{figure}[H]
\centering
\includegraphics[width=0.9\textwidth]{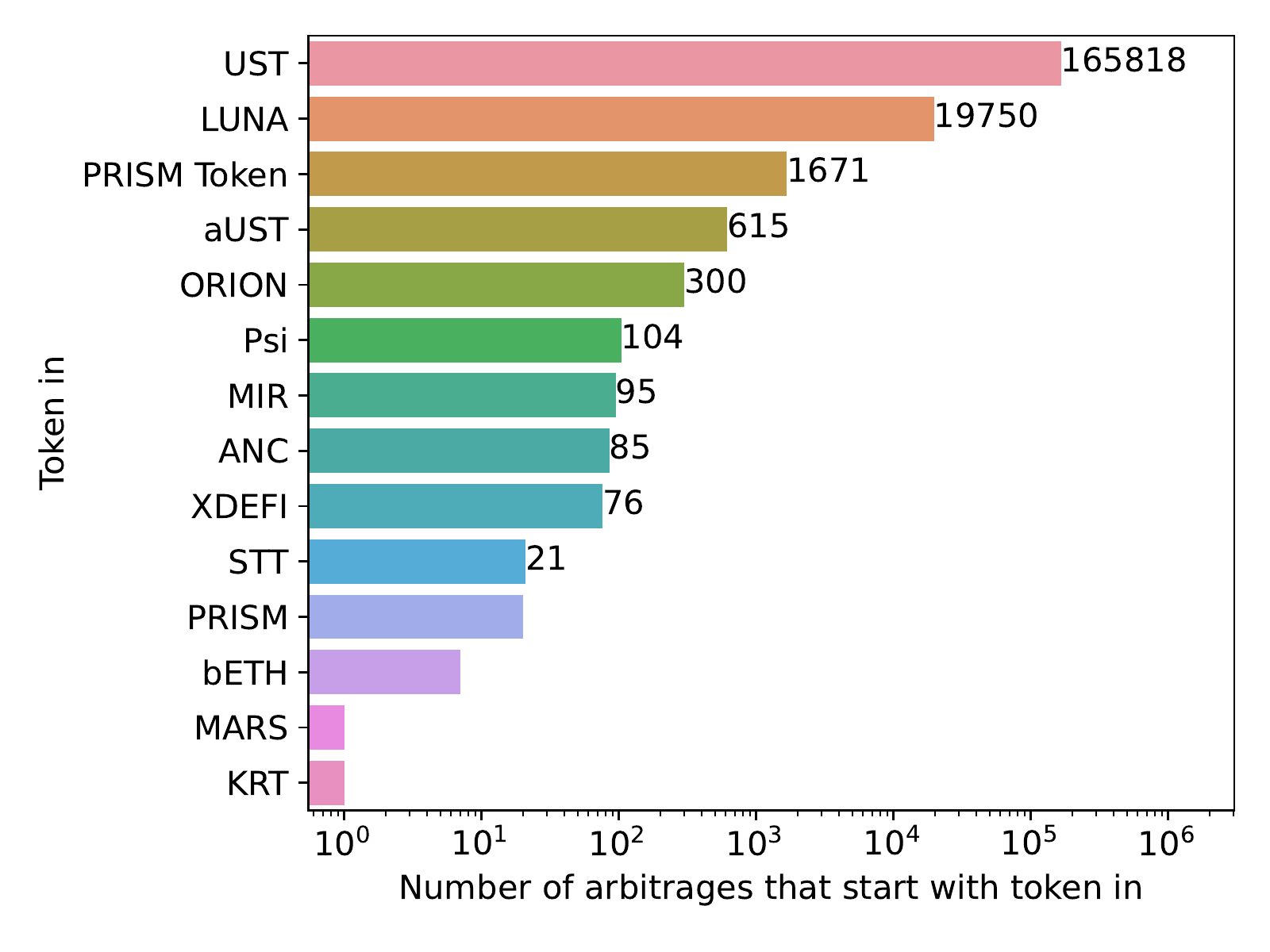}
\caption{\label{fig:fig6_tokenin} Number of arbitrages that start with different tokens. }
\end{figure}

Arbitrages start with different tokens. In our dataset, 87\% (165K) of the arbitrages started with UST. This seems reasonable since UST was the biggest stable coin in Terra Classic before the de-peg event. Following UST, 10\% (19K) of the arbitrages started with LUNA, the native token of Terra Classic. Figure \ref{fig:fig6_tokenin} shows the distribution of the token-in.

\subsubsection{Arbitrage Amount in}

\begin{figure}[H]
\centering
\includegraphics[width=1\textwidth]{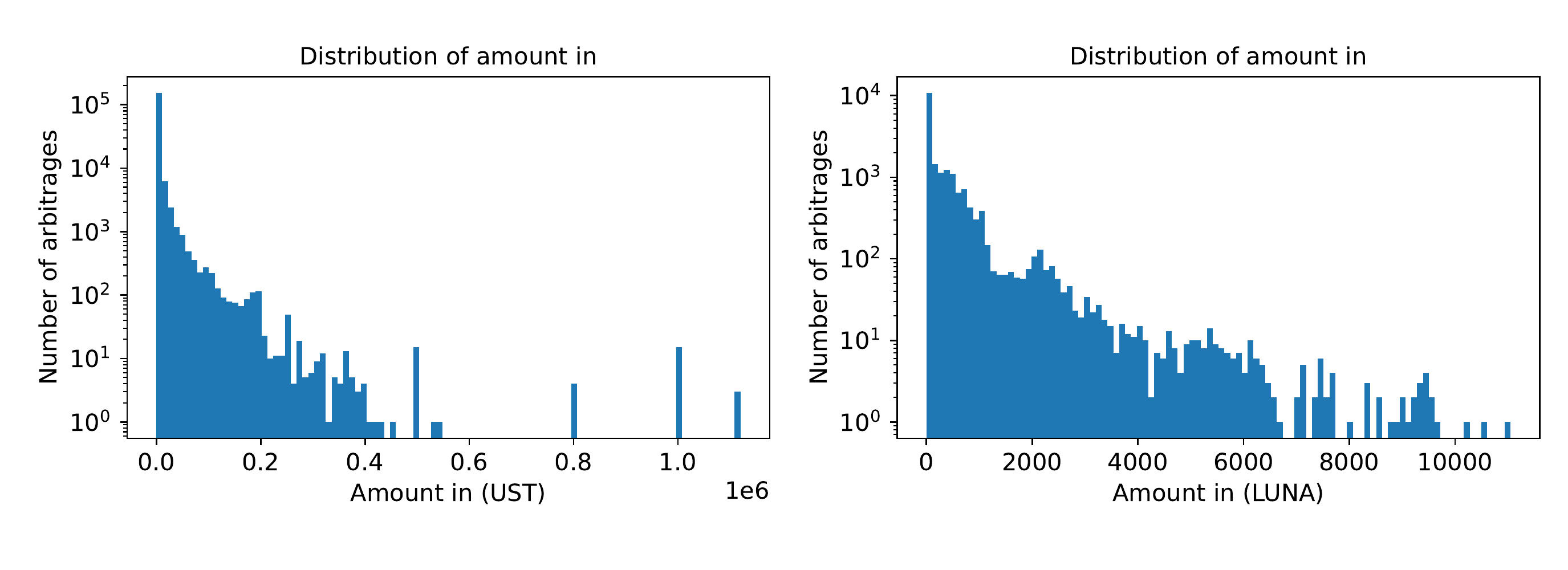}
\caption{\label{fig:fig7_amount_in} The left panel shows the distribution of the number of arbitrages that start with UST. The right panel shows the distribution of the number of arbitrages that starts with LUNA.}
\end{figure}

Arbitrages had a different range of amounts - those started with the UST token shows a median of 998 UST and the 25 percentile is 260 UST. For arbitrage that started with LUNA, the median is 88 LUNA and the 25 percentile is 20 LUNA. Figure \ref{fig:fig7_amount_in} shows these two distributions.

\subsubsection{Path length}

Arbitrage path lengths could be variable. 80\% of the arbitrages use 2 or 3 swaps. Arbitrages with paths greater than five swaps are present in 0.4\% of the arbitrages. Increasing the length of the cycles also increases the cost of gas which reduces profit. So it is reasonable to see the length of the cycles is mostly 2 or 3.

\begin{figure}[H]
\centering
\includegraphics[width=1\textwidth]{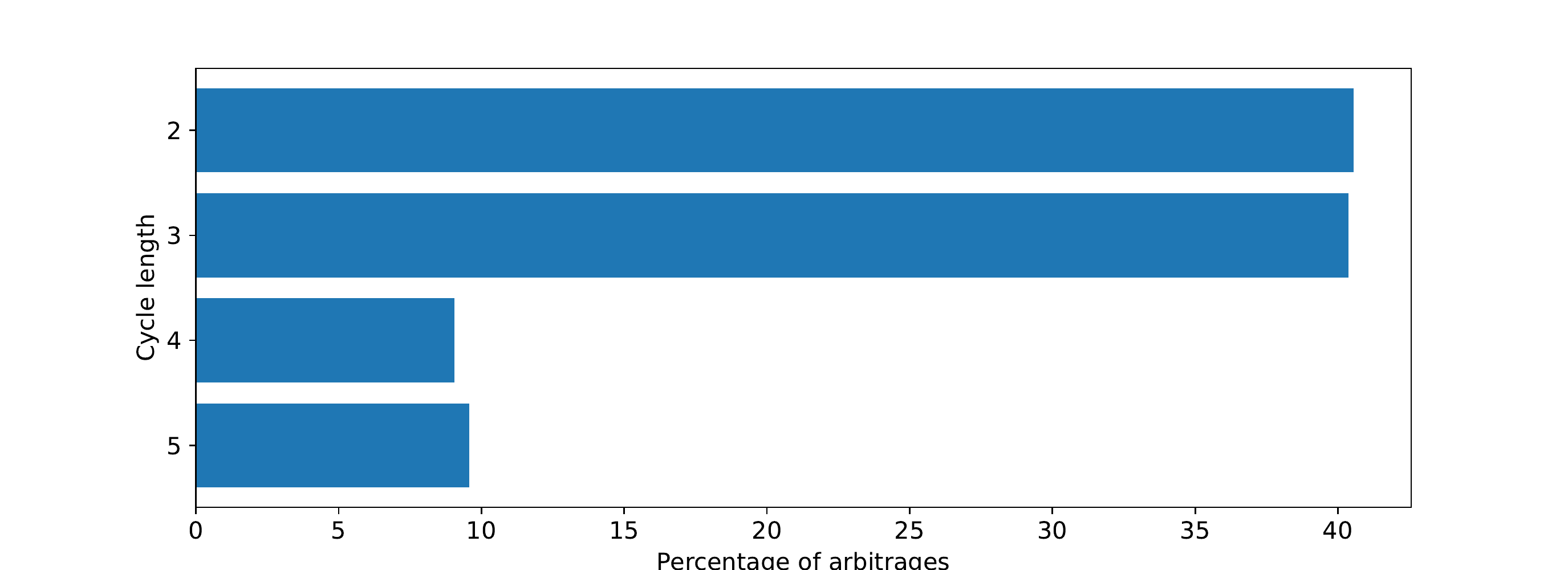}
\caption{\label{fig:fig8_path_len} Percentage of arbitrages according to the cycle length.
}
\end{figure}

\subsubsection{Pairs}
Arbitrages used 300 different pairs/pools. However, the 10 most used pairs cover 38\% of the swaps; the 50 most used pairs cover 82\% and the 100 most used pairs cover 95\%. The top 10 most used pairs are shown in Table \ref{tab:tab3_most_used_pairs}.

\begin{table}
\centering
\begin{tabular}{|l|l|l|l|}
\hline
Pair & Rate & Token & Token \\\hline
terra1m6ywlgn6wrjuagcmmezzz2a029gtldhey5k552 & 0.07316 & UST & LUNA \\\hline
terra1tndcaqxkpc5ce9qee5ggqf430mr2z3pefe5wj6 & 0.07297 & UST & LUNA  \\\hline
terra106a00unep7pvwvcck4wylt4fffjhgkf9a0u6eu & 0.04285 & UST & LOOPC  \\\hline
terra1jxazgm67et0ce260kvrpfv50acuushpjsz2y0p & 0.03370 & bLUNA & LUNA  \\\hline
terra163pkeeuwxzr0yhndf8xd2jprm9hrtk59xf7nqf & 0.03260 & UST & PSI  \\\hline
terra1cda4adzngjzcn8quvfu2229s8tedl5t306352x & 0.03056 & bLUNA & nLUNA \\\hline
terra1c0afrdc5253tkp5wt7rxhuj42xwyf2lcre0s7c & 0.02925 & UST & bETH \\\hline
terra1sgu6yca6yjk0a34l86u6ju4apjcd6refwuhgzv & 0.02562 & UST & LUNA \\\hline
terra1v5ct2tuhfqd0tf8z0wwengh4fg77kaczgf6gtx & 0.02375 & UST & PSI \\\hline
terra1j66jatn3k50hjtg2xemnjm8s7y8dws9xqa5y8w & 0.02339 & bLUNA & LUNA \\\hline

\end{tabular}
\caption{\label{tab:tab3_most_used_pairs}Top 10 most used pairs}
\end{table}

It is worth mentioning that some of the top 10 pairs allow swapping the same pair of tokens. This duplication is because different DEXes offer to swap the same pair of tokens.

\subsection{Profit}

The distribution of arbitrages by searchers is very irregular. Some searchers had good performance while others did not get high profits. 78\% of searchers had ten or more successful arbitrages, 59\% had 100 or more successful arbitrages, and only 35\% had more than 1000 successful arbitrages. 

As mentioned before, the token-in and the token of the profit can vary. Since 97\% of arbitrages used UST or LUNA to start the arbitrage, in this section, we focus only on the arbitrages that used these tokens as token-in.

For those using UST as a token-in, the total profit was 16.24M UST and for LUNA it was 20,569. The value of LUNA can be volatile, unlike stablecoins. In the time range of our dataset, the LUNA price ranged from 35 to 100 UST. Using 100 UST as the upper bound, we limited the profit per 2M UST above (although the value was likely lower).

One way to compare searchers' performance is to use the profit rate. We defined the profit rate as the percentage of profit out of the initial value of the arbitrage - a ratio between profit and the starting value of the arbitrage. For example, if an arbitration starts with 105 UST and 115 UST is obtained, the profit rate is (115-105)/105=0.095. Taking this normalised value into account, 99\% of arbitrages have a profit rate of less than 1. Besides the 99\% of the sample, there were arbitrages with outlier values. For example, this transaction $7ad339227b991c402efaa4c8e02cd7af3c76c5d820d1234cbb759f5f9968aa1e$ \footnote{\href{https://facuzeta.github.io/frp/dashboard/7ad339227b991c402efaa4c8e02cd7af3c76c5d820d1234cbb759f5f9968aa1e/}{https://facuzeta.github.io/frp/dashboard/7ad339227b991c402efaa4c8e02cd7af3c76c5d820d1234cbb759f5f9968aa1e/}}  reached a profit rate higher than 2000.

\begin{figure}[H]
\centering
\includegraphics[width=1\textwidth]{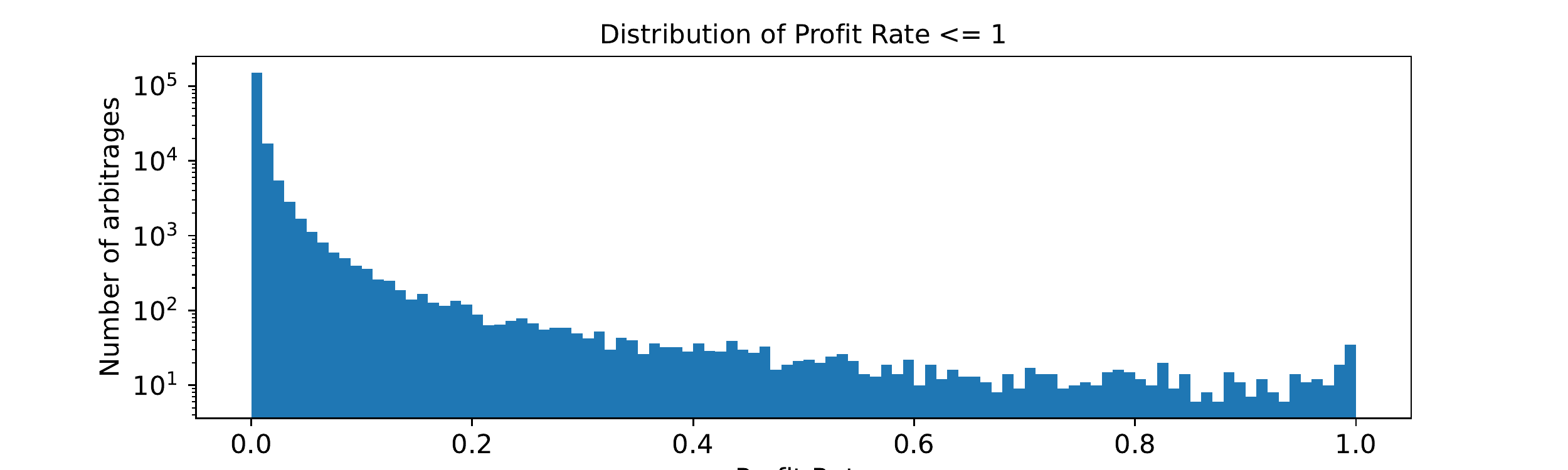}
\caption{\label{fig:fig10_profit_distrubtion}  Distribution of Profit Rate for arbitrages with Profit rate less or equal to 1.
}
\end{figure}

Discarding profit rate values greater than 1, Figure \ref{fig:fig10_profit_distrubtion} shows the distribution of this measure.

Grouping by searchers, we saw that the profit is not uniformly distributed. The three searchers with the highest profit cover more than 75\% of the MEV.

Various characteristics can explain why some searchers have more profit than others. For example, the number of arbitrages they conducted, the number of instances they used to run the bots, and the number of contracts they used, among others.

For arbitrages beginning with UST, the correlation between the profit and the number of arbitrages was positive and statistically significant (Pearson correlation rho=0.8987, p-value$<10^{-18}$). They also were significant and positive correlations, the number of different contracts against the number of arbitrages (rho=0.6945, p-value$<10^{-7}$) and the number of senders against the number of arbitrages (rho=0.5615, p-value$<10^{-5}$). It is relevant to clarify that there was also a significant positive correlation between the number of senders and the number of arbitrages. And also between the number of contracts and the number of arbitrages. Therefore, the previous correlation might only be the product of this phenomenon.

For arbitrages beginning with LUNA, the correlation between the profit and the number of arbitrages is positive and significant (Pearson correlation rho=0.8551, p-value=0.0016). Correlations are also significant and positive in the number of different contracts versus the number of arbitrages (rho=0.6903, p-value=0.0271); and the number of senders versus the number of arbitrages (rho=0.6792, p-value=0.0307). In this case, there is a significant positive correlation between the number of arbitrages and the number of senders. However, there is no correlation between the number of different contracts and the number of arbitrages.

This information is not enough to explain everything. For example whether using more contracts promotes a higher profit. This question can not be answered because profit could also be affected by the number of arbitrages.

To identify which factors contribute to higher profits, it is useful to have a history of the arbitrage attempts that did not work. In the next section, we analyse this factor.

\subsection{Failed arbitrages}

Similar to other blockchains, arbitrages that fail or revert in Terra Classic happened for various reasons. It could be due to estimation errors, or other reasons that caused the swaps to not land on chain. But most likely the transactions revert because another transaction won the opportunity and the profit was already extracted.

In the section \hyperref[sec:inference-of-arbitrage-in-failed-transactions]{Inference of arbitrage in failed transactions}, we introduced a way to estimate reverted arbitrages. The number of arbitrages that reverted was higher than those succeeded. There were 188,564 successful arbitrages and 670,258 failed one. In other words, for each successful arbitrage, there were 3.55 that failed on average.

Considering only executing message-type transactions, successful arbitrages represented 5\% of these transactions and failed arbitrages represented 19\%.

This difference generates negative effects for the network because although they do not increase the value of gas like other chains, they fill the blocks with transactions that revert and increase the size of the mempool, resulting in the network congestion. Analysing it by blocks, the number of failed arbitrages correlates with the number of successful arbitrages (rho = 0.4551, p$<10^{-100}$). The amount of gas used for successful arbitrages was 445 billion, and 1357 billion for failed arbitrages, which is approximately three times of gas used for successful arbitrages.

In the following subsection, we use success rate to help understand the different aspects that determine searchers' profits.

\subsection{Success rate}

We defined the success rate of a searcher as the number of successful arbitrage divided by its own total arbitrage. This measure quantified the searcher with different information. Then, we assess whether there is a correlation between the success rate and the total profit using Spearman correlation. We filter out those searchers with very few successful arbitrages through experimenting with different minimum successful arbitrages as a threshold. In all cases, we obtain a negative correlation between the success rate and profit. This result means that searchers with lower successful-rate had more profit. Table \ref{tab:tab4_success_rate} summarises the correlations.

\begin{table}
\centering
\begin{tabular}{|p{35mm}|p{35mm}|p{35mm}|l|l|}
\hline
Threshold Min  Successful arbitrages & Percentile that threshold represented & Number of Searchers  after threshold & Rho & P-value \\ \hline
10 & 23 & 40 & -0.4869 & 0.0014 \\ \hline
50 & 38 & 32 & -0.5821 & 0.0005 \\ \hline
100 & 44 & 29 & -0.6084 & 0.0005 \\ \hline
250 & 46 & 28 & -0.6349 & 0.0003 \\ \hline
500 & 53 & 24 & -0.5983 & 0.0020 \\ \hline
750 & 55 & 23 & -0.5652 & 0.0049 \\ \hline
\end{tabular}
\caption{\label{tab:tab4_success_rate} Result of correlations for different thresholds}
\end{table}

\begin{figure}[H]
\centering
\includegraphics[width=0.9\textwidth]{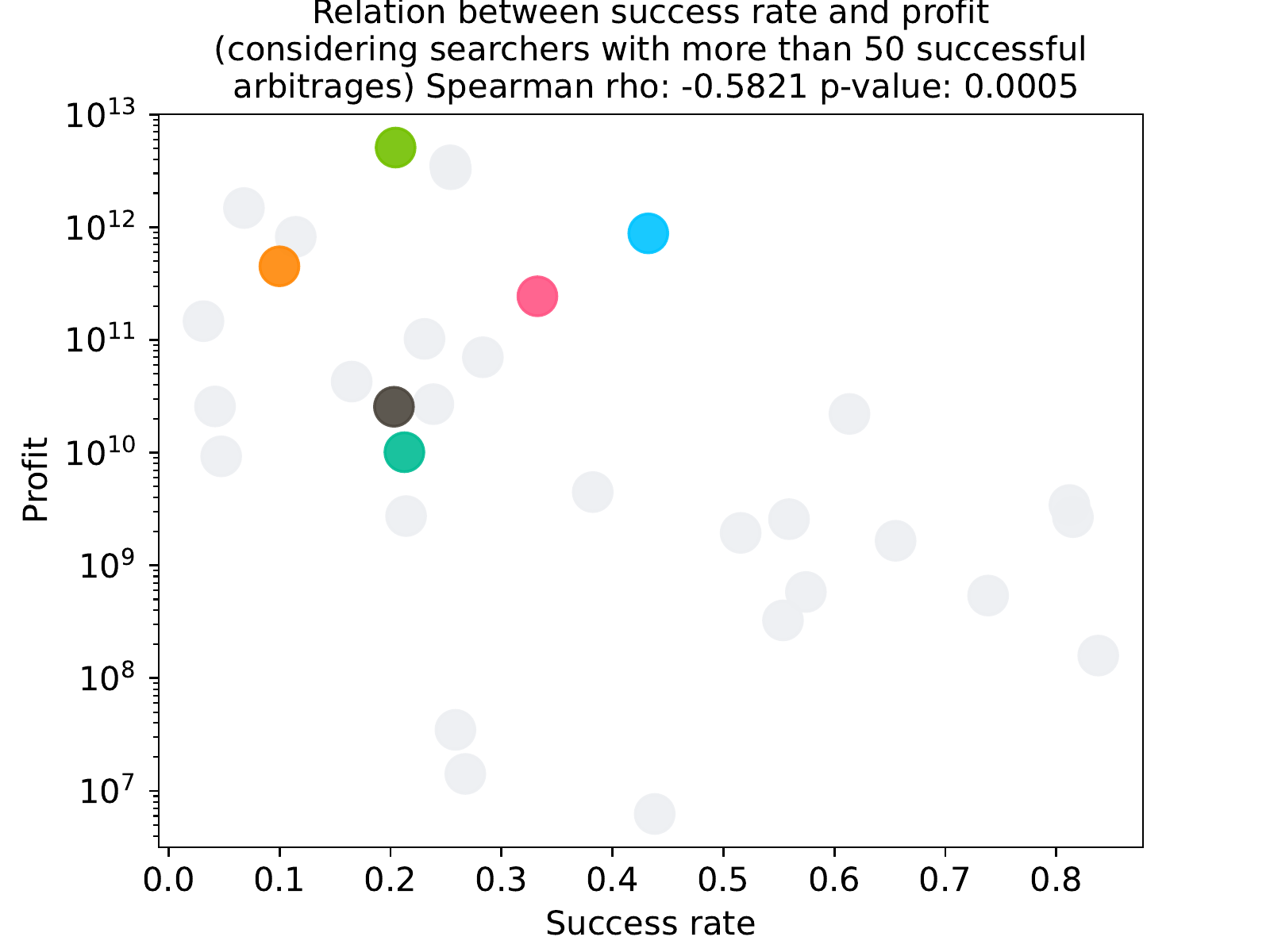}
\caption{\label{fig:fig11_scatter_success_rate_profit} Correlation between success rate and profit by searchers. The colour code follows the one presented in Figure \ref{fig:fig3_graph}. }
\end{figure}

Figure \ref{fig:fig11_scatter_success_rate_profit} shows as an example the scatter plot between the successful rate and the profit for threshold=50 successful arbitrages.

We have already established a negative correlation between profit and the success rate. Those searchers who have a high success rate are not the ones who have a higher profit. The hypothesis behind this phenomenon is that searchers that have many transactions with reverted arbitrages are running bots in multiple instances. They probably have several instances running the same code in different virtual machines. Every instance responds with the same attempt to arbitrage, but one arrives before the others, causing the second and the following to revert.

Multiple running instances could be an effect of the lack of variable gas prices for this type of blockchain. The only mechanism that searchers have to arrive earlier is by reducing the response time. Searchers achieved this by having many instances competing to receive the transaction that generated the opportunity earlier. In the following subsection, we test this hypothesis.

\subsection{Searchers with multiple running instances}

It was impossible to determine if a searcher was running on multiple instances using exclusively on-chain data. However, we can address this problem by studying the reverted transactions and trying to understand, in each case, if they were reverted because another transaction from the same searcher with the same execution message arrived first. If we have many of these transactions, we can assume that this is an effect of the same searcher computing the opportunity to generate a transaction and send it from many instances that are running concurrently.

To test this hypothesis, we built the rate repeated transaction index. This rate measures the times a searcher has repeated the same transaction (in terms of the same execute message) in the same block on average. A high value of this measure indicates that searchers have more than one transaction with the same executed message in the same block. For searchers with at least 50 successful arbitrages, the mean and standard deviation of this measure were 1.3125$\pm$0.5574.

\begin{figure}[H]
\centering
\includegraphics[width=0.9\textwidth]{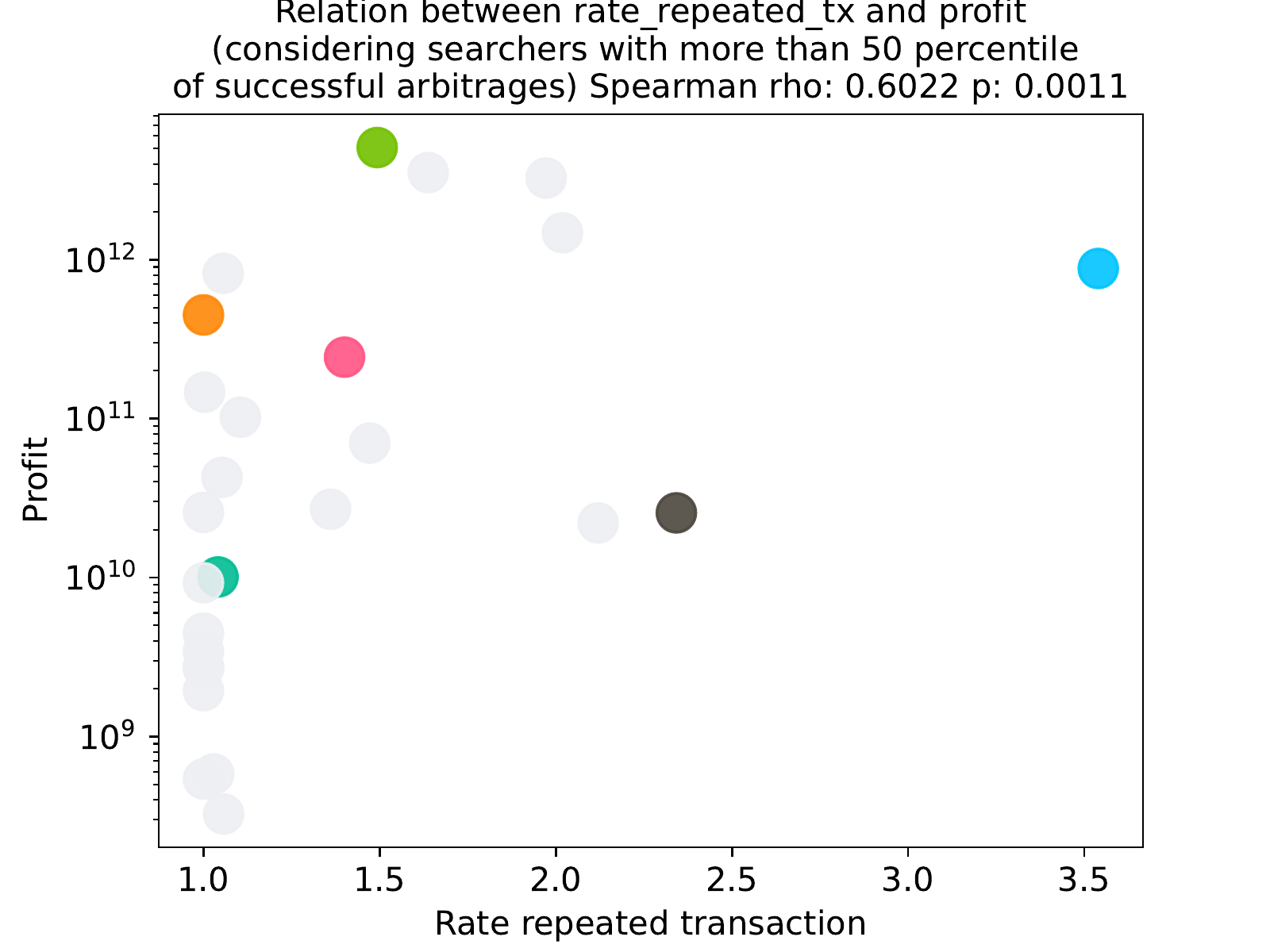}
\caption{\label{fig:fig12_scatter_rate_repeated_tx_profit}  Correlation between repeated transaction rate  and profit by searchers. The colour code follows the one presented in Figure \ref{fig:fig3_graph}.
}
\end{figure}

We also studied the correlation between the rate of repeated transactions and the profit. We found a positive and significant correlation (Spearman correlation rho=0.6022, p-value=0.0011). Figure \ref{fig:fig12_scatter_rate_repeated_tx_profit} shows the scatter plot of this correlation.

This positive and significant correlation explained that searchers that, on average, sent more transactions with the same execute message per block, obtained a higher profit. This is probably explained by the concurrent execution of instances running the same code.

In the next section, we look at the timing and how running multiple instances can make bots see transactions earlier.

\subsection{Time analysis}

In the previous section, we showed that bots with higher profits had higher repeated transaction rate. This effect is probably due to the searchers having more than one instance running the arbitrage bots. The motivation behind this is strictly related to the nature of the fixed gas price blockchains. The only recourse bots have to win an opportunity is to be the fastest.

\begin{figure}[H]
\centering
\includegraphics[width=0.9\textwidth]{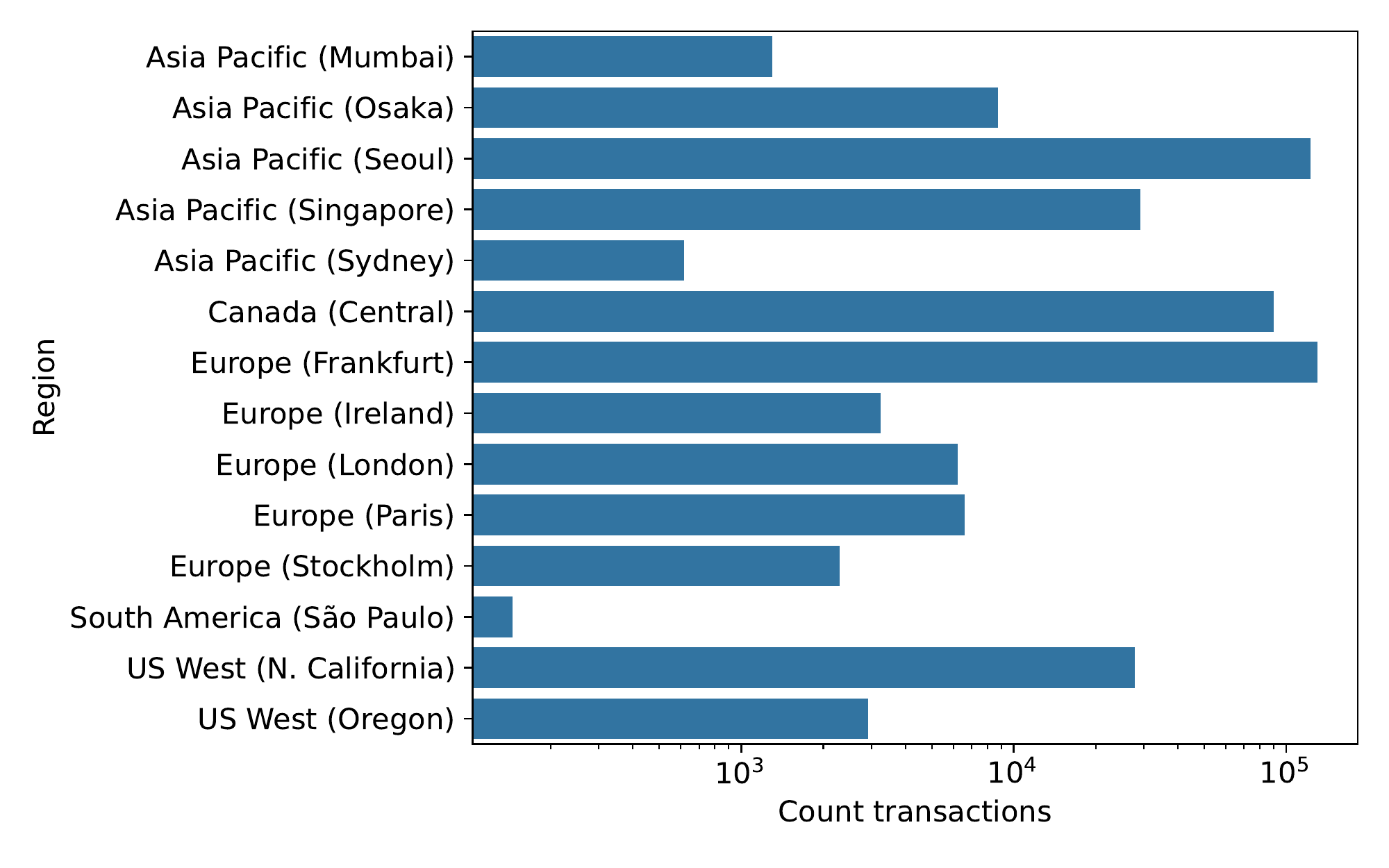}
\caption{\label{fig:fig13_transactions_per_regions} Distribution of times a region saw transactions before others.
}
\end{figure}

We set up the experiment detailed in the  \hyperref[subsubsection:time-analysis]{Time Analysis Section} and measured how long it took for the 84 instances in different geographic locations (running a Terra Classic node) to receive the 431K transactions. For each transaction, we measured the time elapsed relative to the first region that received the transaction. This analysis generated more than 36M different times. Figure \ref{fig:fig13_transactions_per_regions} shows the non-uniform distribution of how many times a region saw a transaction before others.

\begin{figure}[H]
\centering
\includegraphics[width=1\textwidth]{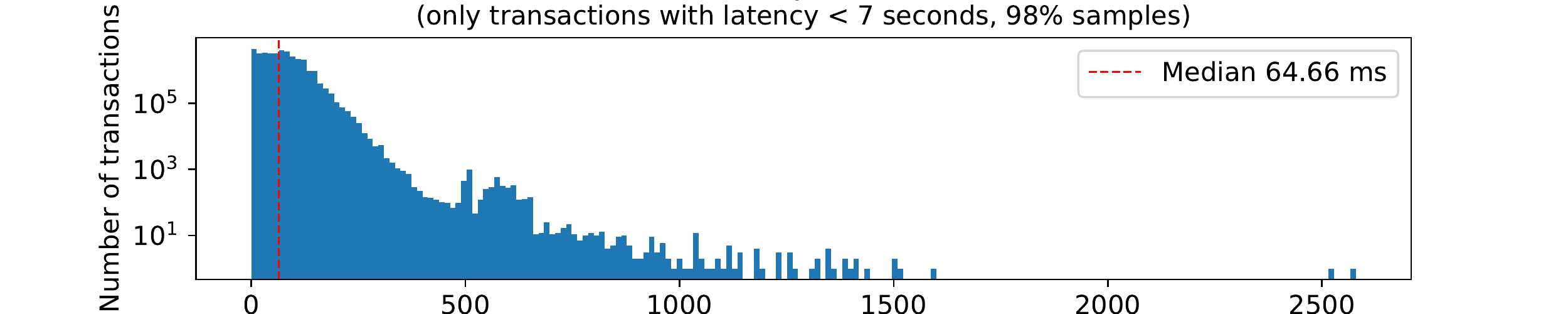}
\caption{\label{fig:fig14_latency_distribution}  Distribution of latency in milliseconds of transactions with latency less than 7 seconds. 
}
\end{figure}

Figure \ref{fig:fig14_latency_distribution} shows the distribution of the latency for transactions that took less than 7 seconds (98\% of the sample).

\begin{figure}[H]
\centering
\includegraphics[width=1\textwidth]{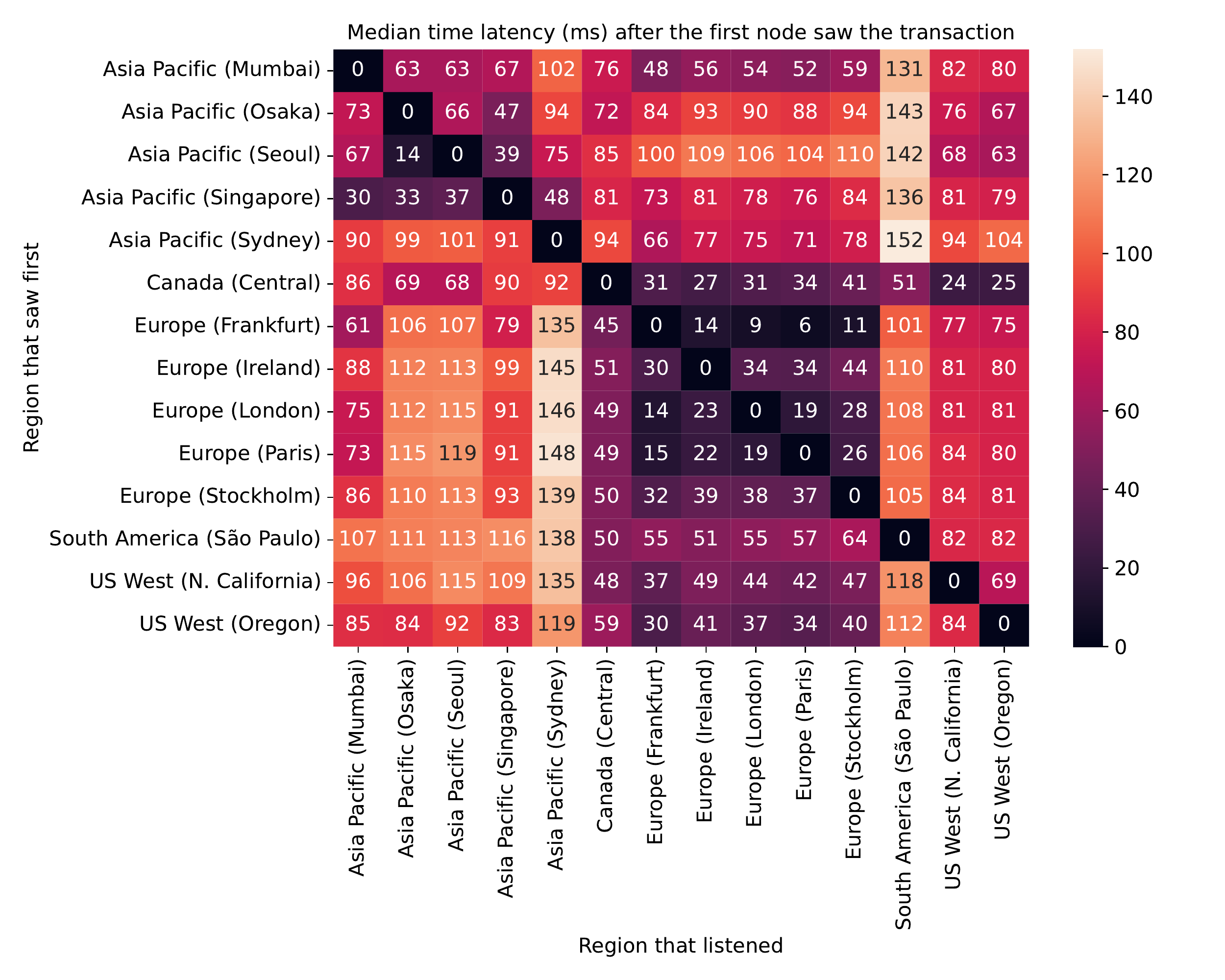}
\caption{\label{fig:fig15_heatmap}  Heat map of median latency in milliseconds. The median time (in milliseconds) that it takes to receive transactions, taking into account the time that elapsed since the first region received the transaction. For example, if Asia Pacific Mumbai received a transaction first, Asia Pacific Osaka waits 63 milliseconds (in median time) and then receives the transaction. Asia Pacific Sydney has to wait for 102 milliseconds, and South America Sao Paulo, 131 milliseconds.
}
\end{figure}

Although the sample distribution of all recorded times is informative, it may be useful to analyse it by region because the effect of geographic proximity may be relevant. We believe that the physically closer the instances are, the faster the transaction arrives. Figure \ref{fig:fig15_heatmap} shows the median time (in milliseconds) that it takes to receive transactions taking into account the time that elapsed since the first region received the transaction. For example, if Asia Pacific Mumbai received a transaction first, Asia Pacific Osaka waits 63 milliseconds (in median time) and then receives the transaction. Asia Pacific Sydney has to wait for 102 milliseconds, and South America Sao Paulo, 131 milliseconds.

\begin{figure}[H]
\centering
\includegraphics[width=1\textwidth]{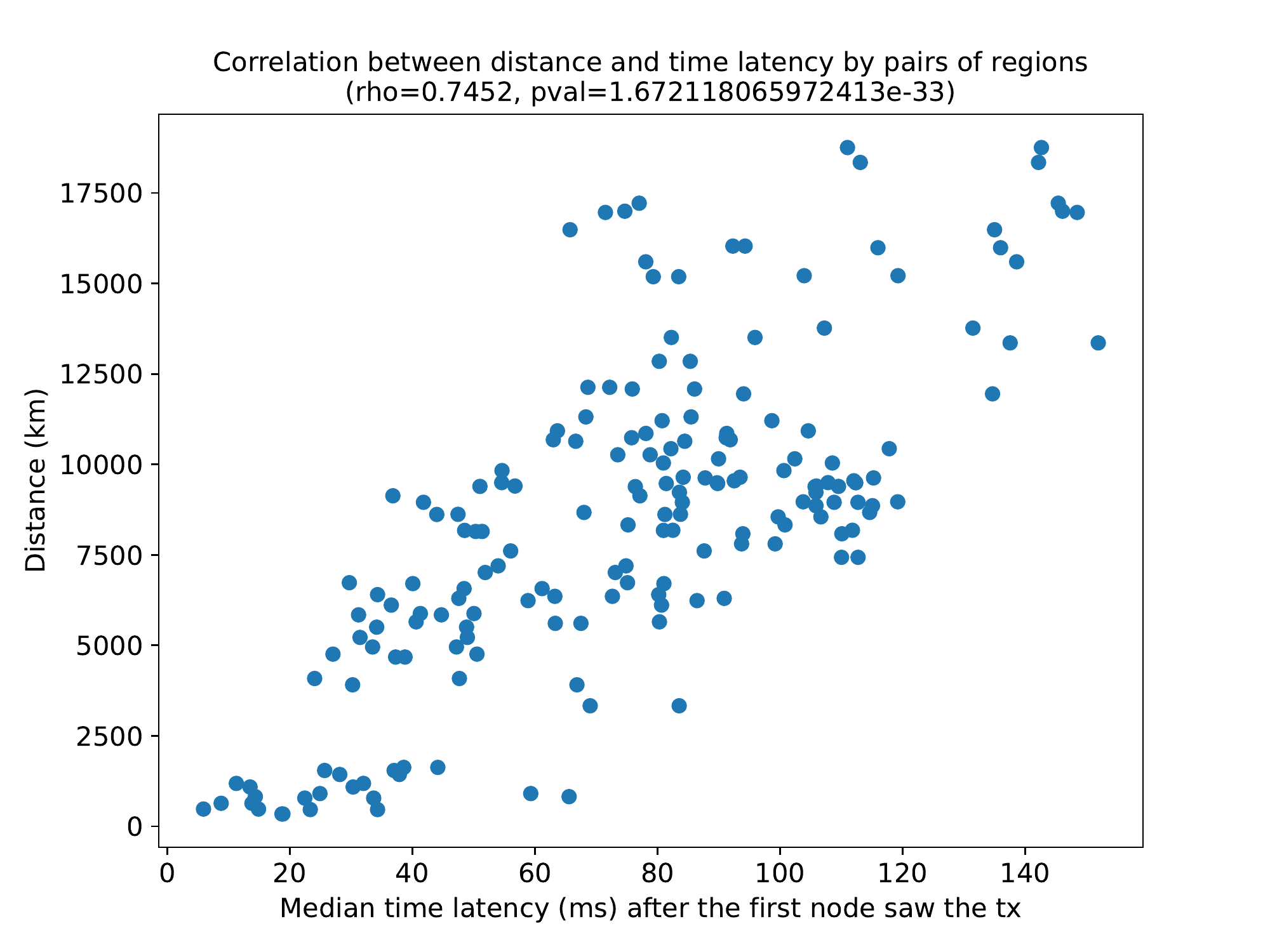}
\caption{\label{fig:fig16_correlation_distance_latency} Correlation between median time latency (in milliseconds) and the distance between regions (in Km).
}
\end{figure}

Figure \ref{fig:fig15_heatmap} shows the expected geographic grouping. We performed the following procedure to quantify the effect. We calculated all pairs of regions and then measured the correlation between each pair of median times with the distance in kilometres between cities of the data centres. We found a positive and significant Pearson correlation (rho=0.7452 and p-value=1.6721e- 33). Figure \ref{fig:fig16_correlation_distance_latency} shows this effect.

We have already established that the time to receive a transaction is related to geographical distances and these values range from 30 to 100 milliseconds in most cases. These characteristics implied that the time in which a bot could see a transaction if it was not in the same region where it originated, was relevant and probably up to an order of magnitude larger than what it may take to compute a solution for a possible arbitrage. Therefore, if another searcher had a bot in the same region where the transaction that generates the arbitrage opportunity originated, it had an advantage of at least 30 milliseconds. This difference suggests that it was relevant for searchers to have many instances running in different regions of the world so that they were always close to the node that first received the transaction that created the opportunity. It is important to note that the time synchronisation error in different instances is less than 10 milliseconds.

\section{Discussion}

In our work, we studied a Terra Classic as an example of a blockchain with a fixed gas price. We use Terra as a case study because of its enormous data volume and the number of decentralised exchanges it has.

We have four specific objectives: Specific Objective 1: Study the characteristics of arbitrages in Terra Classic in the defined period. 
Specific Objective 2: Understanding which strategies improve searchers' profit.
Specific Objective 3: Understanding the timing-related characteristics of the arbitrage.
Specific Objective 4: Create a dashboard to help with the analysis. 

Regarding SO1: In the studied time period, we found more than 188 thousand arbitrages distributed in time proportional to the network load. Typically, they started and ended with two tokens: UST and LUNA. The number of tokens in the arbitrages is very different and the length of the arbitrage cycles is shorter than expected. 80\% of the arbitrages had only 2 or 3 swap and they're mostly from the most exploited pairs - UST and LUNA. We also found some extremely profitable arbitrages, with profit rate greater than 2000. This phenomenon could be due to volatile movements in the token prices at distressed times, breaking the synchronisation of two markets in an outrageous way.

Beyond the individual quantification of the arbitrages, we created a strategy to identify searchers. This task allowed us to make a more precise analysis. We found that a small number of searchers contributed to almost all the profit.

We then proposed a mechanism to identify failed arbitrages which is much larger than the number of successful arbitrages. The ratio was 3.55 failed arbitrages for each successful one. This rate can have negative consequences on the network. Although the gas price cannot be increased, validators have to create larger blocks which increase computation time for all nodes. Also, the amount of data transmitted was larger, congesting and increasing network costs without much real benefits. This could result in transactions staying in the mempool for a long time.

Regarding Specific Objective 2: We understood that several characteristics correlate with high profits. We found that searchers with more complexity (more senders and more contracts) had a higher profit. This complexity could indicate that searchers who constantly changed their strategies with different contracts and frequent improvements eventually generated more profits.

Furthermore, we found that searchers with a lower success rate had a higher profit. At first, this result could sound contradictory. But the fact that searchers sent simultaneous transactions with the same message could mean that searchers ran many instances from various places and they tried to respond with the shortest possible delay. To measure this, we created the rate repeated transition index that allowed us to quantify and corroborate this hypothesis.

Regarding Specific Objective 3: We carried out an observational experiment where we recorded transactions in several instances. After reviewing more than 400K transactions, we found that the physical distance explained the latency in receiving transactions. That means a node has to wait a time proportional to its distance to the source node until it receives the transaction. 
 
Regarding Specific Objective 4: we condensed all this analysis into an interactive platform that allows users to explore the data in a new way. We believe that making these tools available promotes the testing of new hypotheses. The platform also allows for data download which could help the community with further research.

This research study quantifies and characterizes the arbitrage opportunities in Terra Classic before the de-peg event. We believe it is one of the first to address the effects of arbitrages on a fixed gas price blockchain.

In some papers \cite{ko2022analysis}, authors propose that MEV on blockchains like Ethereum is a crypto gas war because bots play a game of finding the right gas price. In the case of blockchains with fixed price gas, we believe that bots are competing in a different type of war: the bot latency war.

\subsection{Limitations}
Our work studied the effects of fixed gas prices on a single blockchain during a particular time. We found features that maximize profit for searchers but these features are not necessarily the same for all blockchains of this type. For future studies, we could test these hypotheses in other blockchains to understand if the same features apply.

Regarding the time experiment, we believe that two improvements could contribute to more robust results. First, our experiment was purely observational. We could study the timing results by sending transactions ourselves in order to have a more precise understanding of the location of the transaction. Secondly, although we control the synchronisation among the different nodes with an error of less than 10 milliseconds, it is possible to reduce this error further and to make our findings more accurate.

\section*{Acknowledgement}
We thank Dr Enzo Tagliazucchi for providing the infrastructure to run the analysis of this work.  We thank the Flashbot team for supporting this project.

\section*{Funding}
This research was supported by the \href{https://github.com/flashbots/mev-research}{Flashbot Research Program}.

\bibliographystyle{unsrt}
\bibliography{sample}

\begin{thebibliography}{10}

\bibitem{qin2022quantifying}
Kaihua Qin, Liyi Zhou, and Arthur Gervais.
\newblock Quantifying blockchain extractable value: How dark is the forest?
\newblock In {\em 2022 IEEE Symposium on Security and Privacy (SP)}, pages
  198--214. IEEE, 2022.

\bibitem{weintraub2022flash}
Ben Weintraub, Christof~Ferreira Torres, Cristina Nita-Rotaru, and Radu State.
\newblock A flash (bot) in the pan: measuring maximal extractable value in
  private pools.
\newblock In {\em Proceedings of the 22nd ACM Internet Measurement Conference},
  pages 458--471, 2022.

\bibitem{flashbotsfrontrunning2022}
Alex Obadia.
\newblock {Flashbots: Frontrunning the MEV Crisis}.
\newblock
  \url{https://medium.com/flashbots/frontrunning-the-mev-crisis-40629a613752},
  2022.

\bibitem{daian2020flash}
Philip Daian, Steven Goldfeder, Tyler Kell, Yunqi Li, Xueyuan Zhao, Iddo
  Bentov, Lorenz Breidenbach, and Ari Juels.
\newblock Flash boys 2.0: Frontrunning in decentralized exchanges, miner
  extractable value, and consensus instability.
\newblock In {\em 2020 IEEE Symposium on Security and Privacy (SP)}, pages
  910--927. IEEE, 2020.

\bibitem{wang2022cyclic}
Ye~Wang, Yan Chen, Haotian Wu, Liyi Zhou, Shuiguang Deng, and Roger
  Wattenhofer.
\newblock Cyclic arbitrage in decentralized exchanges.
\newblock In {\em Companion Proceedings of the Web Conference 2022}, pages
  12--19, 2022.

\bibitem{makarov2020trading}
Igor Makarov and Antoinette Schoar.
\newblock Trading and arbitrage in cryptocurrency markets.
\newblock {\em Journal of Financial Economics}, 135(2):293--319, 2020.

\bibitem{jin2022detecting}
Hai Jin, Chenchen Li, Jiang Xiao, Teng Zhang, Xiaohai Dai, and Bo~Li.
\newblock Detecting arbitrage on ethereum through feature fusion and
  positive-unlabeled learning.
\newblock {\em IEEE Journal on Selected Areas in Communications},
  40(12):3660--3671, 2022.

\bibitem{hansson2022arbitrage}
Magnus Hansson.
\newblock Arbitrage in crypto markets: An analysis of primary ethereum
  blockchain data.
\newblock {\em Available at SSRN 4278272}, 2022.

\bibitem{zhou2021just}
Liyi Zhou, Kaihua Qin, Antoine Cully, Benjamin Livshits, and Arthur Gervais.
\newblock On the just-in-time discovery of profit-generating transactions in
  defi protocols.
\newblock In {\em 2021 IEEE Symposium on Security and Privacy (SP)}, pages
  919--936. IEEE, 2021.

\bibitem{whyyourblockchianneedsandmevsolution}
Flashbots.
\newblock {Why your blockchain needs an MEV solution}.
\newblock \url{https://www.youtube.com/watch?v=sYFuFLe9kp0}, 2022.

\bibitem{abriefsurveyofmev}
Alex Obadia.
\newblock {A brief Survey of MEV on Ethereum, BSC, Avalanche and Polygon in
  2021 }.
\newblock \url{https://www.youtube.com/watch?v=OYE9uAf_v18}, 2022.

\bibitem{haas2017bringing}
Andreas Haas, Andreas Rossberg, Derek~L Schuff, Ben~L Titzer, Michael Holman,
  Dan Gohman, Luke Wagner, Alon Zakai, and JF~Bastien.
\newblock Bringing the web up to speed with webassembly.
\newblock In {\em Proceedings of the 38th ACM SIGPLAN Conference on Programming
  Language Design and Implementation}, pages 185--200, 2017.

\bibitem{forcier2008python}
Jeff Forcier, Paul Bissex, and Wesley~J Chun.
\newblock {\em Python web development with Django}.
\newblock Addison-Wesley Professional, 2008.

\bibitem{ko2022analysis}
Kyungchan Ko, Taeyeol Jeong, Jongsoo Woo, and James Won-Ki Hong.
\newblock An analysis of crypto gas wars in ethereum.
\newblock In {\em 2022 23rd Asia-Pacific Network Operations and Management
  Symposium (APNOMS)}, pages 1--6. IEEE, 2022.

\end{thebibliography}

\end{document}